\documentstyle[psfig,11pt]{nature}

\def\la{\ifmmode\stackrel{<}{_{\sim}}\else$\stackrel{<}{_{\sim}}$\fi} 
\def\ga{\ifmmode\stackrel{>}{_{\sim}}\else$\stackrel{>}{_{\sim}}$\fi} 

\title{The central image of a gravitationally lensed quasar}

\author{\large
Joshua N.\ Winn\affiliation[1]{Harvard-Smithsonian Center for
Astrophysics, Cambridge, MA 02138, USA},
David Rusin$^{1,}$\affiliation[2]{Dept.\ of Physics and Astronomy,
University of Pennsylvania, Philadelphia, PA 19104, USA},
Christopher S.\ Kochanek$^{1,}$\affiliation[3]{Dept.\ of
Astronomy, The Ohio State University, Columbus, OH 43210, USA}
}

\dates{26~September~2003}{2~December~2003}
\headertitle{Central image of a gravitational lens}
\mainauthor{Winn, Rusin, \& Kochanek}

\begin{document}

\summary{ A galaxy can act as a gravitational lens, producing multiple
images of a background object. Theory predicts there should be an odd
number of images\cite{dr80,b81} but, paradoxically, almost all
observed lenses have 2 or 4 images. The missing image should be faint
and appear near the galaxy's center. These ``central images'' have
long been sought as probes of galactic cores too distant to resolve
with ordinary observations.\cite{nsc86,wn93,rm01,eh02,k03} There are
five candidates, but in one case the third image is not necessarily a
central image,\cite{mkk01,l02a,l02b} and in the others, the central
component might be a foreground source rather than a lensed
image.\cite{g83,ch93,f99,w02,wrk03} Here we report the most secure
identification of a central image, based on radio observations of
PMN~J1632--0033, one of the latter candidates. Lens models
incorporating the central image show that the mass of the lens
galaxy's central black hole is $<$2$\times 10^8$~solar masses
($M_{\odot}$), and the galaxy's surface density at the location of the
central image is $>$20,000$M_{\odot}$ per square parsec, in agreement
with expectations based on observations of galaxies hundreds of times
closer to the Earth. } \maketitle

Central images are often called ``odd'' images, but the key property
that makes them interesting is not their odd number, but rather their
proximity to the center of the lens galaxy. To emphasize the
distinction, we note that there can be an odd number of images without
any central images, if the gravitational field has a large
quadrupole\cite{l02a} or if there are multiple lens
galaxies\cite{w03,kw03}.

The absence of central images is a problem dating to the earliest days
of galaxy lens observations, and many solutions have been
proposed.\cite{nsc86,nbn84,scn85} Most recently, Evans \&
Hunter\cite{eh02} and Keeton\cite{k03} argued that the absence of
central images at current detection limits is no longer surprising,
given recent {\it Hubble Space Telescope} observations of nearby
galaxies.\cite{f97} Those observations showed that the distribution of
stars near the centers of massive elliptical galaxies (the predominant
type of lens galaxy) is highly concentrated. Because the central image
flux depends inversely on the square of the surface density, the
concentrated density profiles should cause central images to be very
faint (or even to have zero flux, if the density is singular), and
detection limits will need to be improved by a factor of 10--50 before
central images are commonly seen. In the meantime, the most favorable
systems are radio-loud double quasars with large flux ratios. Radio
waves are not extinguished by dust or overpowered by light in the lens
galaxy, and asymmetric doubles produce the brightest possible central
image for a given mass distribution.

One such system is PMN~J1632--0033.\cite{w02} It has two images (A and
B) of a radio-loud quasar at redshift 3.42, with a flux ratio of
13. The lens redshift is unknown; our working assumption is $z_L=1.0$,
as estimated by requiring the galaxy's photometry and mass to be
consistent with the fundamental plane of elliptical
galaxies.\cite{w02} There is also a faint radio component (C) with a
position and flux appropriate for a central image. However, as in a
few other systems,\cite{g83,ch93,f99} the possibility could not be
excluded that C is an active galactic nucleus (AGN) in the lens galaxy
rather than a third quasar image.\cite{wrk03} The simplest test is to
compare the continuum radio spectra of the components. Because lensing
preserves the frequency of photons, lensed images have the same
spectrum (in the absence of differential propagation effects) whereas
there is no reason why a foreground AGN would have the same spectrum
as a background source.

Previously we attempted this comparison at frequencies from 1.7 to
15~GHz, finding that C was fainter at low frequency than expected for
a third image.\cite{wrk03} However, the discrepancy was limited to a
single measurement at the lowest frequency, where radio propagation
effects are strongest. A central image might be affected more than
other images by scintillation or absorption, due to its passage
through the dense galactic center. Thus we could make no firm
conclusion before obtaining data at higher frequencies, where
propagation effects (scaling characteristically as $\nu^{-2}$) are
negligible.

We have now extended our measurements to 22 and 43~GHz, and obtained
additional data at 8 and 15~GHz, using the Very Large Array (VLA).
The high-frequency spectrum of the central component agrees well with
those of the bright quasar images. For $\nu>1.7$~GHz, the logarithmic
slopes of flux density ratio {\it vs.} frequency (which should be zero
for lensed images) are $0.00\pm 0.04$ for B/A and $-0.02\pm 0.07$ for
C/A. This is powerful evidence that C is a third quasar image.

The evidence that C is not only a {\it third}\, image, but also a
long-sought {\it central}\, image, is its proximity to the center of
the lens galaxy ($\la$30~milli-arcseconds), and its faintness (0.41\%
the flux density of A). This sets PMN~J1632--0033 apart from the only
other three-image system known, APM~08279+5255, in which the lens
galaxy has not been detected, and the fluxes of all three images are
of the same order of magnitude. This leaves open the possibility that
the third image in that system is not a central image, but rather a
``naked-cusp'' image due to a highly flattened mass
distribution.\cite{mkk01,l02a}

For PMN~J1632--0033, the central image properties can be used to
constrain the core structure of the lens galaxy. In general this
requires detailed modeling, for which we refer the reader to Ref.\ 15,
in which some consequences of the central-image hypothesis were worked
out before the status of C was clarified. Here we quote one result,
and elaborate upon it to include the effect of a central black
hole. If the mass distribution is taken to be a spherical power law,
$\rho(r) \propto r^{-\beta}$, embedded in an external shear field to
account for non-sphericity, then the data require $\beta=1.91\pm 0.02$
(2$\sigma$ confidence), only slightly shallower than an isothermal
($\beta=2$) mass distribution. The uncertainty is far smaller for this
system than for typical two-image or four-image systems, because of
the sensitive dependence of the central-image magnification on the
exponent of the central cusp.

If the central mass were too large, then either a bright fourth image
would be produced, or the central image would be de-magnified out of
existence.\cite{nsc86,eh02,k03,mwk01} This fact can be used to derive
an upper limit on the mass of any central black hole. We added a
central point mass to the power-law model described above, finding
that the limits on $\beta$ are barely affected and $M_{\rm BH} <
2.0\times 10^8 M_\odot$.

How does this compare to the mass of the black hole that is expected
to reside in this galaxy? Among nearby galaxies, $M_{\rm BH}$ is
correlated with the velocity dispersion of the surrounding stellar
bulge.\cite{fm00,g00} The exact correlation is a matter of debate, but
for present purposes we adopt the correlation of Tremaine {\it et
al.}\cite{t02} Since the mass distribution of the lens galaxy in
PMN~J1632--0033 appears to be nearly isothermal, the A--B angular
separation $\Delta\theta = 1.46$~arc seconds can be used to estimate
$\sigma$ via the relation $\Delta\theta/2 = (4\pi\sigma^2/c^2) (D_{\rm
LS}/D_{\rm S})$, where $D_{\rm LS}$ and $D_{\rm S}$ are the
lens--source and observer--source angular diameter distances.
Assuming a flat $\Omega_{\rm M}=0.3$ cosmology with
$H_0=72$~km~s$^{-1}$~Mpc$^{-1}$ and $z_{\rm L}=1.0$, the results are
$\sigma=224$~km~s$^{-1}$ and $M_{\rm BH} = (2.1 \pm 0.4) \times 10^8
M_{\odot}$. Thus, the black hole cannot be much larger than expected
from the local correlation.

A simple argument can also be used to place a {\it lower} bound on the
core density of the galaxy. Since the density presumably does not
increase with radius, the minimum central density occurs for the case
of a constant-density core, for which the magnification of C is
$\mu_{\rm C} = (1-\kappa_{\rm C})^{-2}$. Here, $\kappa$ is the surface
density, in units of the cosmology- and redshift-dependent critical
surface density for lensing. The near-isothermal models give $\mu_{\rm
C}=0.0082$, from which it follows $\kappa_{\rm C} > 10$. Under the
same cosmological assumptions as above, the surface density is
$>$20,000\,$M_{\odot}$\,pc$^{-2}$ within the radius of C. The radius
of C is not known accurately but is $\la$30~mas, corresponding to
$\la$230 pc, or $\la$0.15$R_{\rm eff}$, where $R_{\rm eff}$ is the
effective radius of the galaxy.

How does this compare to nearby galaxy cores observed with {\it
HST}\,? Among the sample of Faber {\it et al.}\cite{f97} are 38
early-type (E/S0) galaxies with measurements of core surface
brightness profile, $R_{\rm eff}$, and velocity dispersion. We used
their estimates of the mass-to-light ratio (derived from the velocity
dispersion) to convert the surface brightness profile into a surface
density profile. At $R/R_{\rm eff}=0.15$, the surface densities range
from $10^3$ to $10^5$~$M_{\odot}$\,pc$^{-2}$, showing that our result
is consistent with local estimates and astrophysically relevant.

Thus we have derived limits on surface density and black-hole mass of
the lens galaxy (comoving distance $\sim$3 gigaparsecs) that are
consistent with more detailed measurements possible only for nearby
galaxies ($\sim$30 megaparsecs). The lensing constraints have the
advantage of depending directly on mass, rather than light. This
motivates the discovery of additional central-image systems, and
further observations of this system, some of which we describe below.

The most immediate challenges are direct measurement of the lens
redshift and more accurate ($\la$5~mas) measurement of the position of
C relative to the lens center. These data would better pin down the
physical radius within which the limits on surface density apply, and
firm up the correspondence with nearby galaxies.

Measuring correlated variability between C and another image would
provide irrefutable evidence that C is a central image, should doubt
still remain. Because of the different light-travel times, quasar flux
variations should be seen first in A, then $\approx$180 days later in
B, and then $\approx$9 hours later in C. The ratio of the delays would
also further constrain mass models, although the C--B delay
measurement would be difficult due to the faintness of the image and
the necessarily fine time-sampling.

Finally, an additional image\cite{mwk01} is predicted to occur over
the entire allowed range of parameters of our models with central
black holes. In general, the fourth image has $<$10\% the flux of C,
although there are very rare arrangements in which C represents a pair
of merging images. It is separated by $<$20~mas from C, indicating the
need for very-long-baseline interferometry. Currently, the strongest
upper limit on the flux of an additional component is $<$30\% that of
C, based on the 5$\sigma$ limit at 8~GHz.\cite{wrk03} Detection of the
fourth image would be interesting because the separation and
magnification ratio between the central images would provide a
measurement of $M_{\rm BH}$, rather than an upper limit.

\bibliographystyle{nature}
\bibliography{journals}

\clearpage
\newpage

\onecolumn

\begin{acknowledge}
The authors are grateful to S.\ Doeleman and P.\ Schechter for helpful
discussions, and J.\ Bullock for comments on the manuscript. J.N.W.\
acknowledges the support of the National Science Foundation (NSF)
through an Astronomy \& Astrophysics Postdoctoral Fellowship. The VLA
is part of the National Radio Astronomy Observatory, an NSF facility
operated under cooperative agreement by Associated Universities, Inc.
\end{acknowledge}

{\raggedright Correspondence should be addressed to J.N.W.\ (e-mail:
jwinn@cfa.harvard.edu)}.

\vspace{2in}

\begin{table}[hb]
\small
\begin{center}
\begin{tabular}{lccccc}
\hline
\hline
       Date & Frequency & A     & B      & C      & R.M.S.\ noise \\
       (UT) & (GHz)     & (mJy) &  (mJy) &  (mJy) & (mJy beam$^{-1}$) \\ \hline
2003~Jul~23 & 8.46  & 202.13  &  16.49   & 0.84   & 0.033 \\
2003~Aug~02 & 8.46  & 209.82  &  16.65   & 0.74   & 0.040 \\
2003~Aug~30 & 8.46  & 195.19  &  15.86   & 0.78   & 0.042 \\
2003~Jun~10 & 14.94 & 173.27  &  12.50   & 0.72   & 0.16  \\
2003~Jul~01 & 14.94 & 180.17  &  13.87   & 0.57   & 0.13  \\
2003~Jun~12 & 22.46 & 149.54  &  10.78   & 0.62   & 0.076 \\
2003~Jun~24 & 22.46 & 146.95  &  10.81   & 0.48   & 0.061 \\
2003~Jun~26 & 22.46 & 146.42  &  10.78   & 0.66   & 0.066 \\
2003~Jun~15 & 43.34 & 107.25  &   7.85   & 0.38   & 0.077 \\
\hline
\hline
\end{tabular}
\end{center}

\caption{\small { \bf Radio observations of PMN~J1632--0033 with the
Very Large Array.} The array was in the A configuration. The primary
flux density calibrator was 3C~286 and the phase calibrator was
J1658+0741 (except the 8~GHz observation of 2003~Jul~23, for which it
was J1651+0129). Initial calibration was performed with standard AIPS
procedures. Further self-calibration and imaging were performed with
Difmap. Flux densities were determined by fitting a model consisting
of 3 point sources to the visibility data. For the 8~GHz and 15~GHz
data, the relative positions of the components were fixed at the
values determined by previous higher resolution
measurements\cite{w02}. }
\label{tab:radio}
\end{table}

\clearpage
\newpage

\begin{figure}[htpb]

\vspace{0.5in}
\centerline{\psfig{file=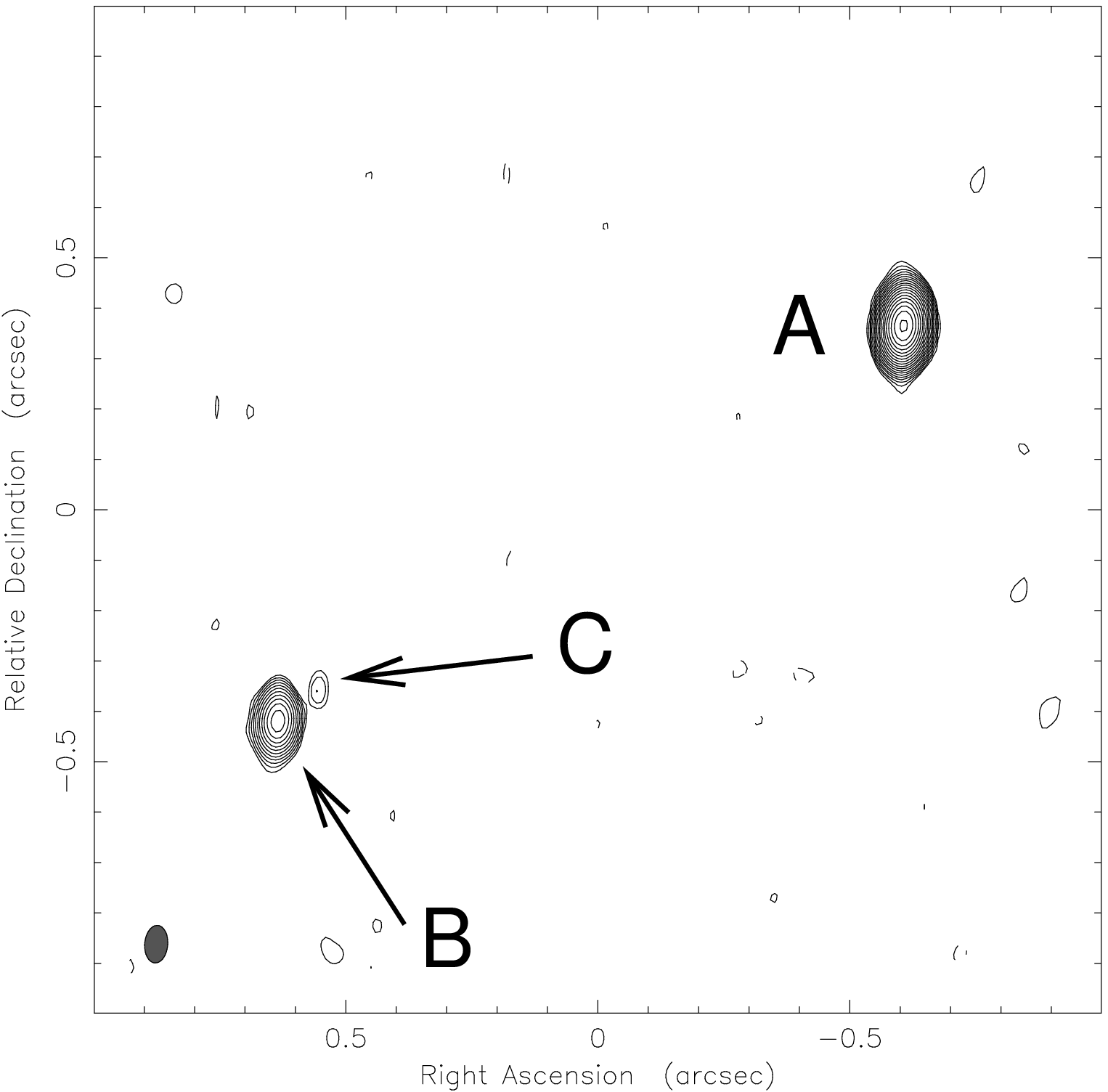,width=5in}}
\vspace{0.2in}
{\small{{\bf{Figure 1.}}
A radio map of the three components of gravitational
lens PMN~J1632--0033. The observations were carried out with the VLA
at 43~GHz, with a total bandwidth of 100~MHz. Standard phase and
amplitude calibration were applied, and the map was created with
natural weighting. The root-mean-squared (RMS) noise level in the
residual map is $\sigma=0.077$~mJy~beam$^{-1}$, as compared to the
theoretical minimum RMS of $0.04$~mJy~beam$^{-1}$. Contour levels
begin at $2.5\sigma$ and increase by powers of $\sqrt{2}$. The
synthesized beam (full width at half-maximum of $0.073\times 0.044$
arc seconds) is illustrated in the lower left corner of the map.}}
\label{fig1}
\end{figure}

\clearpage
\newpage

\begin{figure}[htbp]
\centerline{\psfig{file=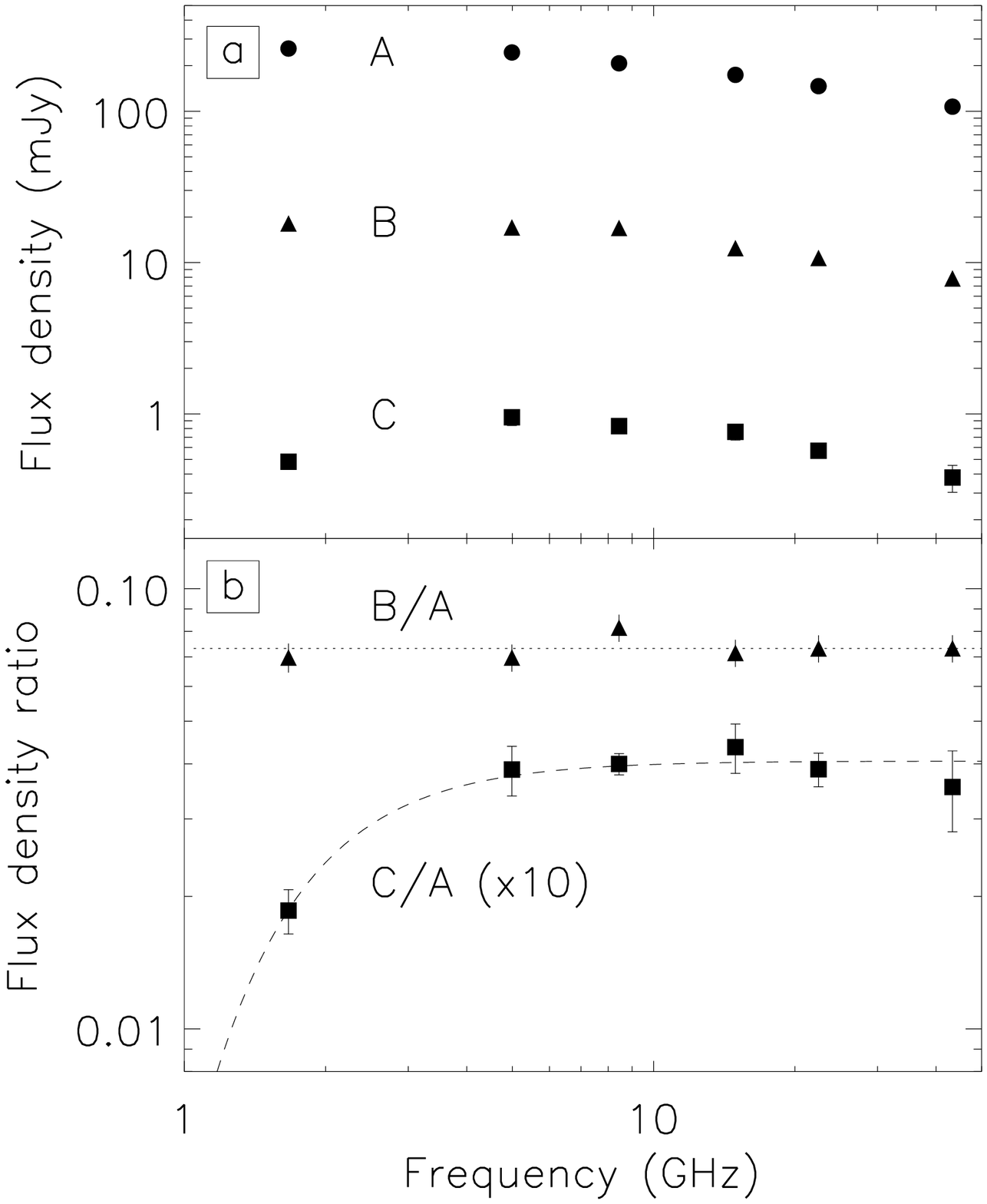,width=4in}}
\end{figure}
{\small{{\bf{Figure 2.}}  The central radio component and the bright
quasar images have similar radio spectra. (a) The flux density of each
component as a function of frequency. (b) The flux density ratios
relative to A. Results for C/A were multiplied by 10 for display
purposes. All the observations at a given frequency from Table~1 and
Ref.\ 15 were combined by averaging the flux density ratios and
adopting the total flux density of the most recent measurement, to
avoid problems due to source variability and inconsistencies in
absolute flux density scale. The only discrepancy between C and A is
at 1.7~GHz, where C is fainter than expected for a lensed image. One
plausible explanation is free-free absorption due to ionized material
$\la$200 parsecs from the lens galaxy nucleus, the approximate
position of C. The dashed line is a 2-parameter fit to the function
$\mu_{\rm CA} e^{-\tau_\nu}$, using a standard
approximation\cite{mh67} for free-free opacity, $\tau_\nu =
0.08235$~$T^{-1.35} \nu^{-2.1} E$, where $\nu$ is the frequency (in
GHz) and $T$ and $E$ are the temperature (in Kelvins) and emission
measure (the integral of $n_e^2$ over the path length $s$, in
cm$^{-6}$pc) of the ionized medium. The required opacity $\tau_\nu =
0.7$ at the lens-frame frequency $\nu\approx 3.4$~GHz could be
produced by circumnuclear material with, for example, $T=6000$ and
$E=10^7$ ($n_e=10^3$~cm$^{-3}$, $s=10$~pc), which seems physically
reasonable. For comparison, the emission measure is larger than in
most individual Galactic H~{\sc ii} regions ($E$$\sim$10$^{5-6}$;
e.g., ref.\ 25), comparable to some ionized clouds around Sgr~A
($E$$\sim$10$^{6-7}$; e.g., ref.\ 26), and smaller than observed
around the nuclei of some radio galaxies ($E$$\sim$10$^{8}$, e.g.,
refs.\ 27,28). The free-free absorption hypothesis and the nature of
the absorbing medium could be tested further with future low-frequency
observations.}}

\end{document}